\documentclass[twocolumn]{aastex631}
\usepackage{amsmath}

\usepackage{CJK}

\begin{document}
\begin{CJK*}{UTF8}{gbsn}

%%%%%%%%%%%%%%%%%%%%%%%%%%%%%%%%%%%%%%%%%%%%%%%%%%%%%
%%%%%%%%%%%%%%%%%%%
%%%    Author List
%%%%%%%%%%%%%%%%%%%
\correspondingauthor{Peng Wang}
\email{pwang@shao.ac.cn}

\title{Observed Anti-parallel Correlation Between Spiral Galaxy and Cosmic Filament Spins}

\author{Hao-da Wang (汪昊达)}
\affil{Shanghai Astronomical Observatory, Nandan Road 80, Shanghai 200030, China}
\affil{University of Chinese Academy of Sciences, Beijing 100049, China}

\author[0000-0003-2504-3835]{Peng Wang* (王鹏)}
\affil{Shanghai Astronomical Observatory, Nandan Road 80, Shanghai 200030, China}

\author[0009-0005-9342-9125]{Min Bao (鲍敏)}
\affil{School of Astronomy and Space Science, Nanjing University, Nanjing 210023, China}
\affil{Key Laboratory of Modern Astronomy and Astrophysics (Nanjing University), Ministry of Education, Nanjing 210023, China}

\author[0000-0003-3226-031X]{Yanmei Chen (陈燕梅)}
\affil{School of Astronomy and Space Science, Nanjing University, Nanjing 210023, China}
\affil{Key Laboratory of Modern Astronomy and Astrophysics (Nanjing University), Ministry of Education, Nanjing 210023, China}

\author[0009-0001-7527-4116]{Xiao-xiao Tang (唐潇潇)}
\affil{Shanghai Astronomical Observatory, Nandan Road 80, Shanghai 200030, China}
\affil{University of Chinese Academy of Sciences, Beijing 100049, China}

\author{Youcai Zhang (张友财)}
\affil{Shanghai Astronomical Observatory, Nandan Road 80, Shanghai 200030, China}

\author[0000-0002-5458-4254]{Xi Kang (康熙)}
\affil{Institute for Astronomy, School of Physics, Zhejiang University, Hangzhou 310027, China}
\affil{Center for Cosmology and Computational Astrophysics, Zhejiang University, Hangzhou 310027, China}

\author{Quan Guo (郭铨)}
\affil{Shanghai Astronomical Observatory, Nandan Road 80, Shanghai 200030, China}

\author[0000-0002-9891-338X]{Ming-Jie Sheng(盛明捷)}
\affil{Department of Astronomy, Xiamen University, Xiamen, Fujian 361005, People's Republic of China}

\author[0000-0001-5277-4882]{Hao-Ran Yu(于浩然)}
\affil{Department of Astronomy, Xiamen University, Xiamen, Fujian 361005, People's Republic of China}

\begin{abstract}

Understanding the origin of galactic angular momentum and its connection to the cosmic web remains a pivotal issue in galaxy formation. Using kinematic data from the MaNGA survey, we investigate the alignment between the spin directions of spiral galaxies and their host cosmic filaments. By incorporating filament spin measurements derived from redshift asymmetry across filament spines, we reveal a mass-dependent anti-parallel correlation: low-mass spiral galaxies ($\log_{10}(M_*/M_\odot) \lesssim 10$) exhibit a statistically significant anti-parallel alignment between their stellar/gas spins and filament spins, while high-mass spirals show no such trend.  Spatial analysis further indicates that high-mass spirals preferentially reside near filament spines, whereas low-mass spirals occupy filament outskirts. These findings extend previous alignment studies that neglected directional spin correlations and provide new insights into how cosmic environments shape galactic angular momentum. The observed anti-parallel trend suggests a critical role for filament spin in regulating the angular momentum acquisition of low-mass spirals. This anti-parallel alignment is significantly enhanced for low-mass spirals residing in dynamically cold filaments, highlighting the importance of filament properties in shaping galaxy spin.

\end{abstract}

\keywords{
    \href{http://astrothesaurus.org/uat/902}{Large-scale structure of the universe (902)};
    \href{http://astrothesaurus.org/uat/330}{Cosmic web (330)};
    \href{http://astrothesaurus.org/uat/2029}{Galaxy environments (2029)};
    %\href{http://astrothesaurus.org/uat/1882}{Astrostatistics (1882)}
    \href{http://astrothesaurus.org/uat/602}{Galaxy kinematics (602)};
    \href{http://astrothesaurus.org/uat/595}{Galaxy formation (595)}
}

\section{Introduction} \label{sec:intro}

Understanding how galaxies acquire their angular momentum remains a fundamental question in cosmology and galaxy formation. According to tidal torque theory  (TTT) \citep{Hoyle1949,1969ApJ...155..393P, 1970Ap......6..320D,1979MNRAS.186..133E,1984ApJ...286...38W, 1987ApJ...319..575B,2007JRSSC..56....1S} and detailed physical studies of TTT \citep{2002MNRAS.332..325P,2002MNRAS.332..339P, 2021MNRAS.502.5528L,2024PASP..136c7001L,2025arXiv250501298L}, galaxies gain angular momentum through gravitational interactions with the surrounding tidal field during their formation. As matter accretes from the cosmic web, these tidal interactions induce rotational motion in proto-galactic structures, shaping their final angular momentum. Consequently, the orientation of a galaxy's angular momentum is expected to correlate with the local tidal field \citep{1993ApJ...418..544V, 1996MNRAS.281...84V,2000ApJ...532L...5L,2007MNRAS.375..489H,2007MNRAS.381...41H,2008LNP...740..335V,2009IJMPD..18..173S,2009MNRAS.398.1742H,2012MNRAS.427.3320C,2024arXiv240716489K}. In particular, within filamentary environments, the elongated morphology of filaments, combined with the coherent flow of matter along their axes, results in a strong alignment between the angular momentum of galaxies and the filament structure.

A mass-dependent spin-filament correlation was first discovered in N-body simulations \citep{2007MNRAS.381...41H,2007ApJ...655L...5A} and is confirmed in Horizon 4$\pi$ N-body simulation \citep{2012MNRAS.427.3320C}. Furthermore, high resolution N-body simulations such as Planck-Millennium $\Lambda$CDM simulation \citep{2018MNRAS.481..414G,2021MNRAS.503.2280G}, hydrodynamical simulations such as Horizon-AGN \citep{2014MNRAS.444.1453D}, EAGLE \citep{2019MNRAS.487.1607G} and Illustris \citep{2018ApJ...866..138W,2023ApJ...954...49Z} have investigated the spin-filament correlation as a function of galaxy properties, including stellar mass, color, and star formation rate. Their results indicate that low-mass, blue galaxies preferentially align their spin vectors with the nearest filaments, whereas high-mass, red galaxies tend to show a perpendicular alignment. This spin-flip phenomenon is also investigated in a theoretical context, explained by the TTT \citep{2015MNRAS.452.3369C,2020OJAp....3E...3N,2021MNRAS.502.5528L}

Observationally, studies based on Sloan Digital Sky Survey (SDSS) data \citep{2013ApJ...775L..42T, 2013MNRAS.428.1827T} have confirmed that the spin-filament correlation depends on galaxy morphology: spiral galaxies tend to align their spins with nearby filaments, whereas elliptical galaxies exhibit a perpendicular alignment between their short axes and the filament direction. Additionally, \citet{2015ApJ...798...17Z} found a mass-dependent correlation, showing that the spin of spiral galaxies weakly aligns with or is perpendicular to the intermediate or minor axis of the local tidal tensor.

Further analysis using the 2MASS Redshift Survey \citep{2012ApJS..199...26H},  \citet{2016MNRAS.457..695P} examined spin alignment with the V-web velocity shear field \citep{2012MNRAS.425.2049H, 2013MNRAS.428.2489L}. Their results revealed a significant perpendicular alignment for elliptical galaxies with the slowest compression axis, while no significant signal was detected for spirals. In contrast, \citet{2019ApJ...876...52K} analyzed data from the MaNGA integral field survey \citep{2015ApJ...798....7B, MaNGA-sample, MaNGA-pipeline} and found no strong evidence for a correlation between galaxy spins and filament orientations, though a potential mass-dependent trend was noted. More recently, \citet{2020MNRAS.491.2864W} observed the flip of galaxy spin and successfully measured the stellar transition mass in the SAMI Galaxy Survey \citep{2021MNRAS.505..991C}. \citet{2022MNRAS.516.3569B} analyzed the correlation of the alignments with the bulge mass of galaxies, and \citet{2023MNRAS.526.1613B} used the SAMI Galaxy Survey to investigate the influence of black hole activity on galaxy spin-filament alignments, demonstrating that active galactic nuclei (AGN) can affect the orientation of gas spins relative to filaments. \citet{2025MNRAS.538.2660B} also investigated the effect of filaments on the amplitude of galaxy spin rather than the orientation using the SAMI Galaxy Survey, revealing the relationship of the distance to filaments, nodes or voids with the amplitude of galaxy spin.

Despite significant progress, previous studies have primarily focused on the alignment of angular momentum orientations while neglecting the directionality of these alignments. Specifically, most analyses consider only the absolute value of the cosine of alignment angles, disregarding the statistical significance of their directional values. However, recent findings by \citet{Wang2021NA, 2025ApJ...982..197T} suggest that filaments themselves exhibit intrinsic spin, introducing asymmetry along the filament axis. Filament spin is also confirmed in simulations \citep{2021MNRAS.506.1059X, 2022PhRvD.105f3540S, 2025ApJ...983..100W}. This implies that studies of galaxy alignment must account not only for angular momentum orientation but also for the actual spin direction of filaments. Determining filament spin directions enables an assessment of whether galaxies preferentially parallel  or anti-parallel with their host filaments.

Accurately measuring the spin direction of galaxies is equally crucial. While some studies infer galaxy orientations from photometric position angles, precise determination of angular momentum directions requires kinematic measurements. In this work, we utilize kinematic data from the MaNGA survey, which employs integral field spectroscopy to derive precise kinematic position angles. The method described in \citet{Wang2021NA} is adopted to determine the spin directions of filaments. By comparing these measurements, we aim to provide observational evidence for the relationship between galaxy angular momentum and the surrounding large-scale structure.

This paper is organized as follows. Section \ref{sec:method} describes the data and methodology. The results are presented in Section \ref{sec:results}, followed by discussions and conclusions in Section \ref{sec:s&d}.

\begin{figure*}[!htp]
    \centering
    \plotthree{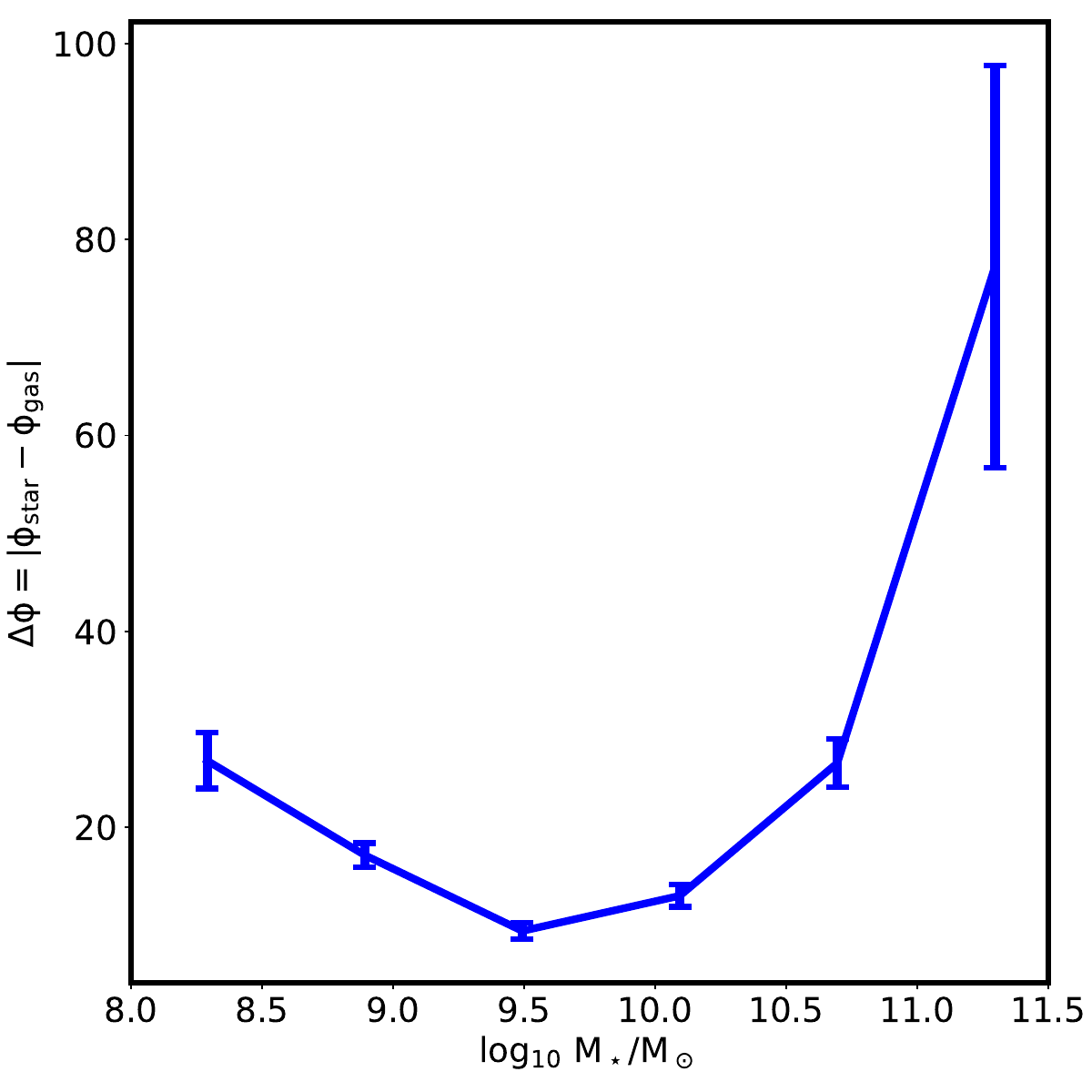}{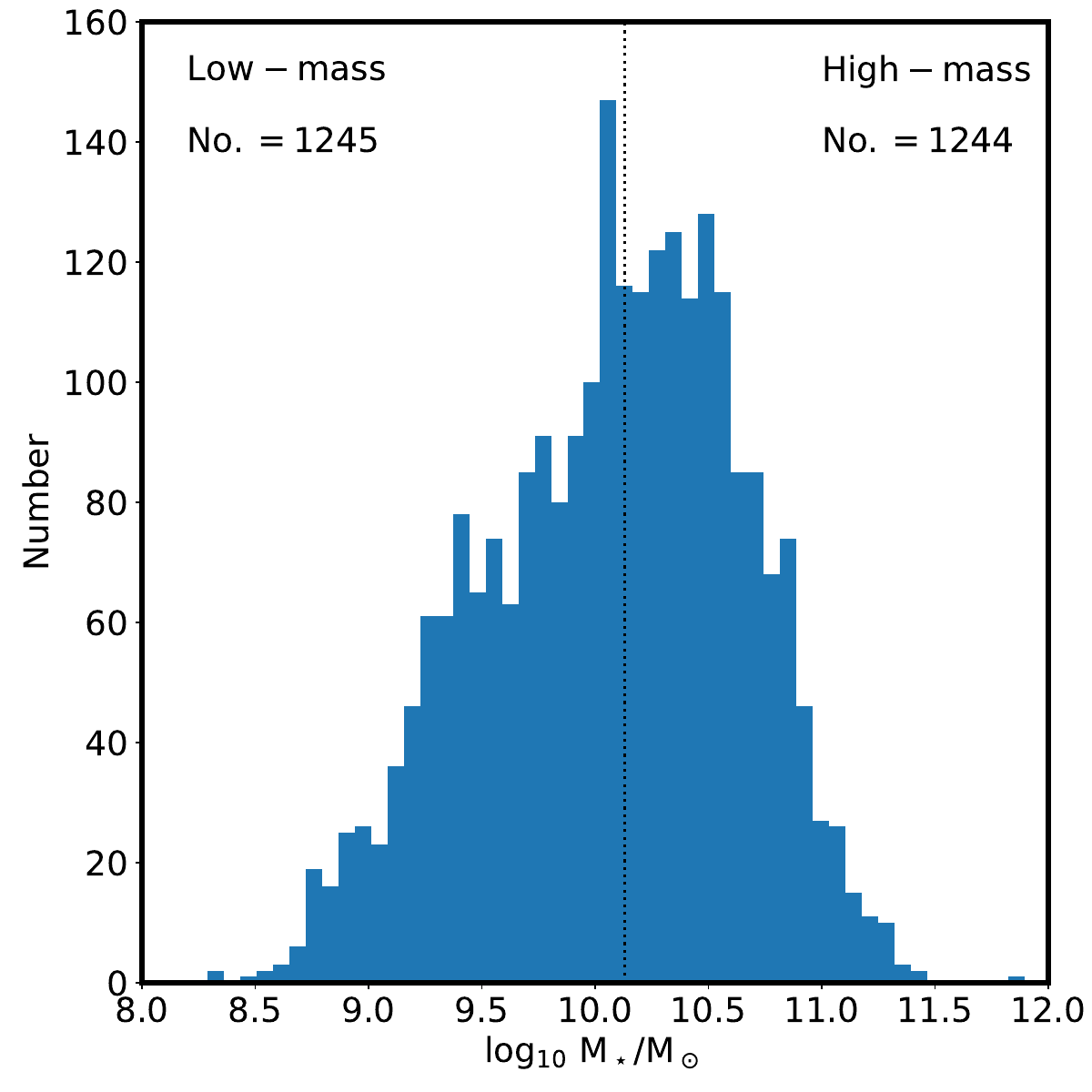}{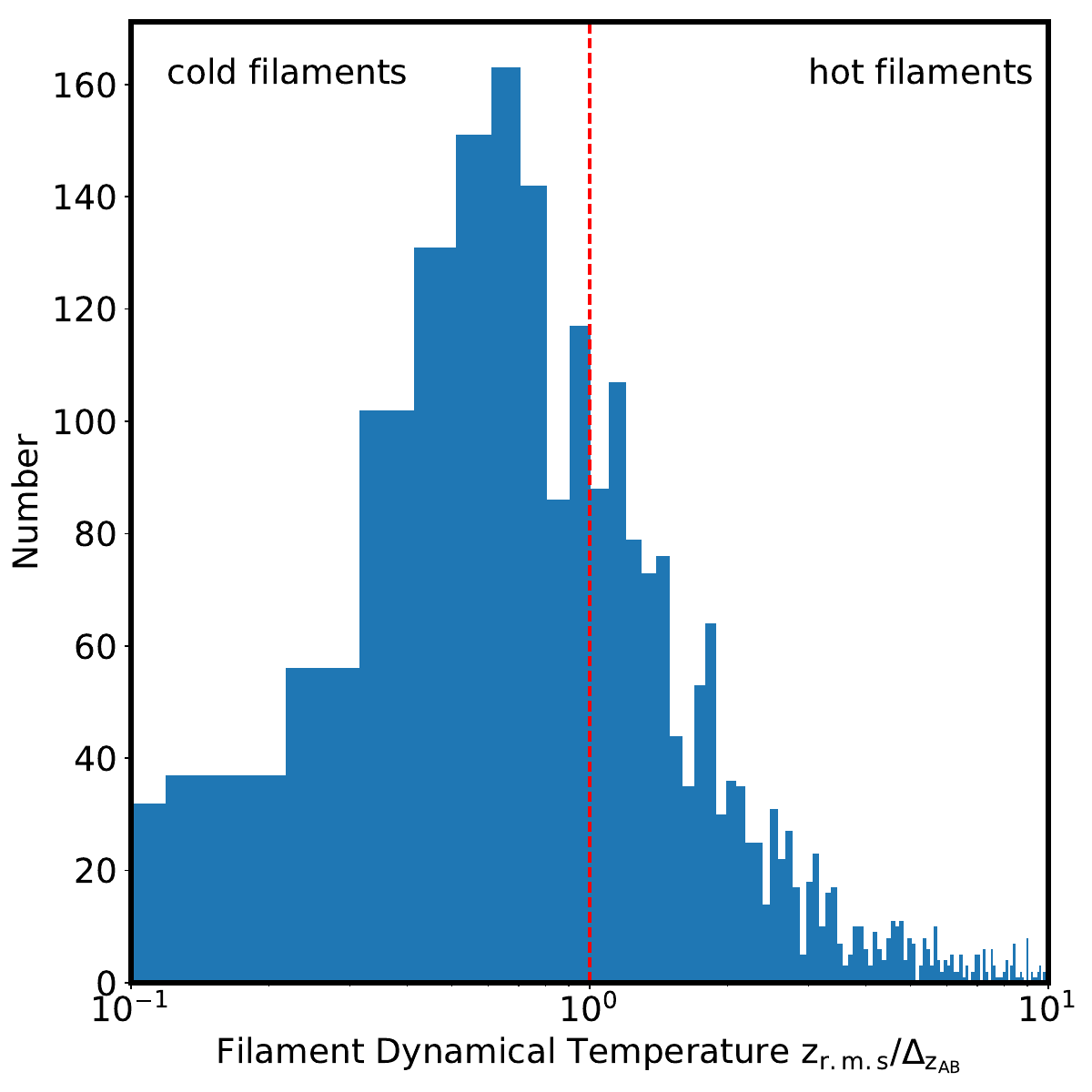}
    \caption{\textbf{Left panel:} the position angle misalignment,  $\rm \Delta\phi \equiv \phi_{star} - \phi_{gas}$, between stellar and gas components as a function of galaxy stellar mass, $\rm \log_{10}(M_\star/M_\odot)$. Blue line with error bars correspond to spiral galaxies. Error bars represent typical $1\sigma$ uncertainties. \textbf{Middle panel:} the number distribution of spiral galaxies as a function of galaxy stellar mass, $\rm \log_{10}(M_\star/M_\odot)$. Galaxies are divided into two sub-samples of almost equal size according to the  $\rm \log_{10}(M_\star/M_\odot) = 10.118$, indicating by the vertical dotted line. \textbf{Right panel:} the distribution of filament dynamical temperature $\rm z_{r.m.s}/\Delta_{Z_{AB}}$ (see section 2.2 for more details). Filaments are divided into either `cold' or `hot' by $\rm z_{r.m.s}/\Delta_{Z_{AB}}=1$, indicating by the vertical red dotted line. 
    }
    \label{fig:f1}
\end{figure*}

\section{Data and Method} \label{sec:method}

\subsection{The MaNGA galaxies}

We selected galaxies from the MaNGA survey (Mapping Nearby Galaxies at Apache Point Observatory; part of SDSS-IV, \citealp{2015ApJ...798....7B,MaNGA-sample}). MaNGA employs 17 filament-bundle integral field units (IFUs) and dual-beam spectrographs mounted on the 2.5m Sloan Telescope (\citealp{2006AJ....131.2332G}) to observe approximately 10,000 low-redshift galaxies. The spectra are processed through the MaNGA Data Analysis Pipeline (DAP; \citealp{MaNGA-pipeline}), and the position angles used in this study, derived from MaNGA's velocity fields, are consistent with PAFIT-based calculations (\citealp{2022MNRAS.515.5081Z}).

Global properties for each MaNGA galaxy, including right ascension (RA), declination (Dec), redshift (z) and minor-to-major axial ratio (b/a), are sourced from the NSA catalog (Blanton et al. 2011), while stellar mass ($M_{\star}$) is derived from the MaNGA DRP catalog (Law et al. 2016). Morphological classifications are obtained from Galaxy Zoo 2 (\citealp{Galaxy-zoo-2}) and Galaxy Zoo DECaLS (\citealp{Galaxy-Zoo-DECaLS}).

Following the methods of \citet{Lee2007, 2012ApJ...744...82V, 2021MNRAS.504.4626K, 2015ApJ...798...17Z}, the spin vector of a galaxy is computed as
\[
\begin{aligned}
S_x &= \cos \alpha \cos \delta \sin \zeta + \cos \zeta (\sin \phi \cos \alpha \sin \delta - \cos \phi \sin \alpha), \\
S_y &= \sin \alpha \cos \delta \sin \zeta + \cos \zeta (\sin \phi \sin \alpha \sin \delta + \cos \phi \cos \alpha), \\
S_z &= \sin \delta \sin \zeta - \cos \zeta \sin \phi \cos \delta,
\end{aligned}
\]
where $\alpha$ and $\delta$ denote the right ascension and declination, respectively, and $\phi$ is the position angle. 
Only galaxies with spaxel-level signal-to-noise ratios (S/N $> 3$) are considered: we fit the kinematic position angles of ionized gas ($\phi_{\mathrm{gas}}$) and stellar ($\phi_{\mathrm{star}}$ ) velocity fields using the Python-based package \texttt{PAFIT}. The velocity field of ionized gas is derived from H$\alpha$ emission line and the stellar component is derived from continuum \citep{2016MNRAS.463..913J,2021MNRAS.504.4626K,2022MNRAS.511.4685X,2022MNRAS.515.5081Z,2025ApJ...982L..29B}. The position angle is defined as a counterclockwise angle between north and a line bisecting the velocity field. We only fit the spaxels with H$\alpha$ signal-to-noise ratio (S/N) higher than 3 for the gas velocity field, and the spaxels with median spectral S/N higher than 3 for the stellar velocity field.
The inclination angle $\zeta$, defined as the angle between the galactic disk and the line of sight, is determined via
\[
\sin^2 \zeta = \frac{(b/a)^2 - f^2}{1 - f^2},
\]
where we adopt \( f = 0.158 \) as suggested by \citet{2021MNRAS.504.4626K}. Tests with various values of \( f \) indicate that the results are not significantly affected.

This study only focus on spiral galaxies, primarily attributed to the observational advantage in determining angular momentum vectors for this morphological class compared to their elliptical. As established in our introductory discussion, while prior investigations (e.g., \citealt{Tempel-2013}) have proposed tentative alignments between spiral galaxies and large-scale filaments, persistent discrepancies remain among published findings (\citealt{2015ApJ...798...17Z}; \citealt{2016MNRAS.457..695P}). These inconsistencies appear fundamentally tied to methodological variations in cosmic web definition and sample selection criteria. Crucially, we emphasize the astrophysical distinction between alignment phenomena (angular momentum vectors parallel to filaments) and anti-parallel configurations: these represent distinct physical phenomena that likely originate from different formation mechanisms within the cosmic structure framework.

Figure~\ref{fig:f1} (left panel) shows the distribution of the angular offset,
\[
\Delta \phi = |\phi_{gas} - \phi_{star}|,
\]
as a function of galaxy stellar mass. The blue curve indicates that at the low-mass end ($\log_{10}(M_\star/M_\odot) \sim 8.0-9.5$) $\Delta\phi$ decreases sharply, reaching a minimum near $\log_{10}(M_\star/M_\odot) \sim 9.5$, indicative of a strong alignment between stellar and gas components. For higher masses, $\Delta \phi$ increases gradually, suggesting a loss of coherent orientation in more massive systems. Studies \citep{2016NatCo...713269C, 2022MNRAS.515.5081Z} have reported observation evidence of galaxies with misalignment between stellar and gas angular momentum. \citet{2025ApJ...982L..29B} also claimed that such misalignment is related to the alignment between galactic spin and filaments. We have tested selecting galaxies with $\Delta \phi$ less than 30 or 60 degrees, and found no significant effect on the conclusions.

\begin{figure}
    \centering
    \plotone{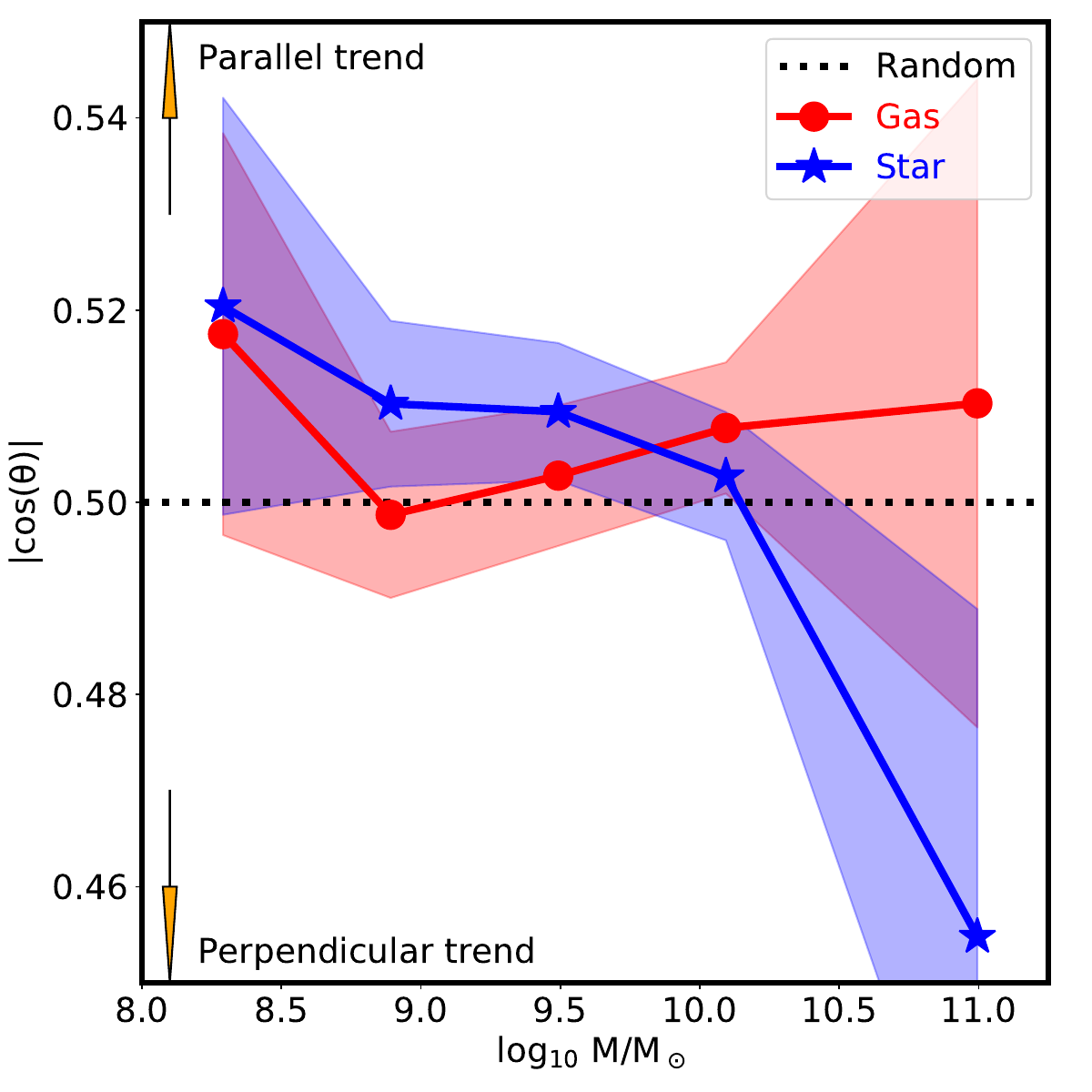}
    \caption{The spiral galaxy spin-filament correlation, $\rm |\cos(\theta)|$, as a function of galaxy stellar mass, $\rm log_{10} M_\star/M_\odot$. The horizontal dotted line ($\rm |\cos(\theta)|=0.5$) indicate the galaxy spin is randomly distributed with respect to the filament orientation. The solid blue line with stars and red line with circles represent the stellar and gas components, respectively, as indicated in the legend. The shaded regions in corresponding colors represent the $1/\sqrt{N}$ statistical uncertainty, where $N$ is the number of galaxies in each bin. Yellow arrowheads denote trends aligned either parallel or perpendicular.}
    \label{fig:f2}
\end{figure}

\subsection{Filament spin}

The filament catalogue used in this work is the same as that used in \citet{2020ApJ...900..129W}, which was based on SDSS data release 12 \citep{Eisenstein2011,SDSS-III}, and filaments are traced by the bisous process \citep{2007JRSSC..56....1S} and \citet{2016A&C....16...17T} according to the spatial distribution of galaxies. We adopt the same method introduced in \citet{Wang2021NA} to determine the spin direction of the filaments. 

In short, we propose that if filaments undergo rotation, a significant velocity component should be present perpendicular to their axis. This velocity component would manifest in opposite directions on either side of the filament's spine, with one side moving away and the other moving toward the observer. In essence, for each filament, we categorize the galaxies into two regions (either A or B). We then calculate the redshift difference ($\rm \Delta z_{AB}$) between two regions, where the region with a higher redshift is interpreted as moving away from the observer, while the region with a lower redshift is approaching. This allows us to determine the filament's spin direction. Details of this method are introduced in \citet{Wang2021NA, 2025ApJ...982..197T, 2025ApJ...983..100W}. The confidence level of spin measurements for individual filament was evaluated. In the right panel of our Figure~\ref{fig:f1}, we present the distribution of dynamical temperatures ($\rm z_{r.m.s}/\Delta_{Z_{AB}}=1$) for the filaments in our sample, where $\rm z_{r.m.s}$ is the root mean square of redshift in a given filament. It is evident that the majority are dynamically cold. According to the findings in Figure 1 of \citet{Wang2021NA}, these cold filament exhibit higher confidence levels.

Based on previous experience in studying the connection between galaxies and filaments, and particularly considering the statistical result from \citet{Wangwei2024}, we select galaxies located within 2 Mpc of the filament spine. It is important to note that the radius of filaments is inherently multiscale and can vary depending on the identification algorithm, as discussed in, for example, \citet{2014MNRAS.441.2923C}. The 2 Mpc scale adopted here does not imply that all filaments have the same radius, but rather serves as a representative value widely used in previous studies to characterize the region where filaments exert significant influence on galaxies. After assigning the galaxy sample to filaments, we obtained 2,489 spiral galaxies for analysis. To facilitate subsequent analysis, we divided the sample into two equally sized subsamples (1,245 low-mass and 1,244 high-mass galaxies) at a stellar mass threshold of $\log_{10}(M_\star/M_\odot) = 10.118$, as shown in the right panel of Figure~\ref{fig:f1}.

\subsection{Galaxy spin - filament spin correlation}
To describe the spin-LSS correlation, we calculate the value of the cosine angle between the spin vector ($\vec{S}$) of a given spiral galaxy and the direction of the spin ($\vec{e}_f$) of the filament.

\[
\cos(\theta) = \frac{\vec{S} \cdot \vec{e}_f}{|\vec{S}| \cdot |\vec{e}_f|}
\]

It is worth emphasizing that in the MaNGA survey, the position angles of galaxies are measured over the full range of $0^{\circ}$ to $360^{\circ}$, allowing the determination of the three-dimensional spin vectors $\vec{S}$ of a galaxy. This represents one of the key advantages of using MaNGA data in our analysis. In case $\rm \cos(\theta)$ is close to 1, we refer to a parallel alignment between the galaxy spin and the filament spin, and for $\rm \cos(\theta)$ close to -1, we call it an anti-parallel between the galaxy spin and the filament spin. \citet{2016arXiv161206873C} had discussed the distribution of $\rm \cos(\theta)$. However, the result was totally symmetric and the different physics nature of parallel and anti-parallel were not taken into consideration.

\section{Results} \label{sec:results}

\begin{figure*}[!htp]
    \centering
    \plottwo{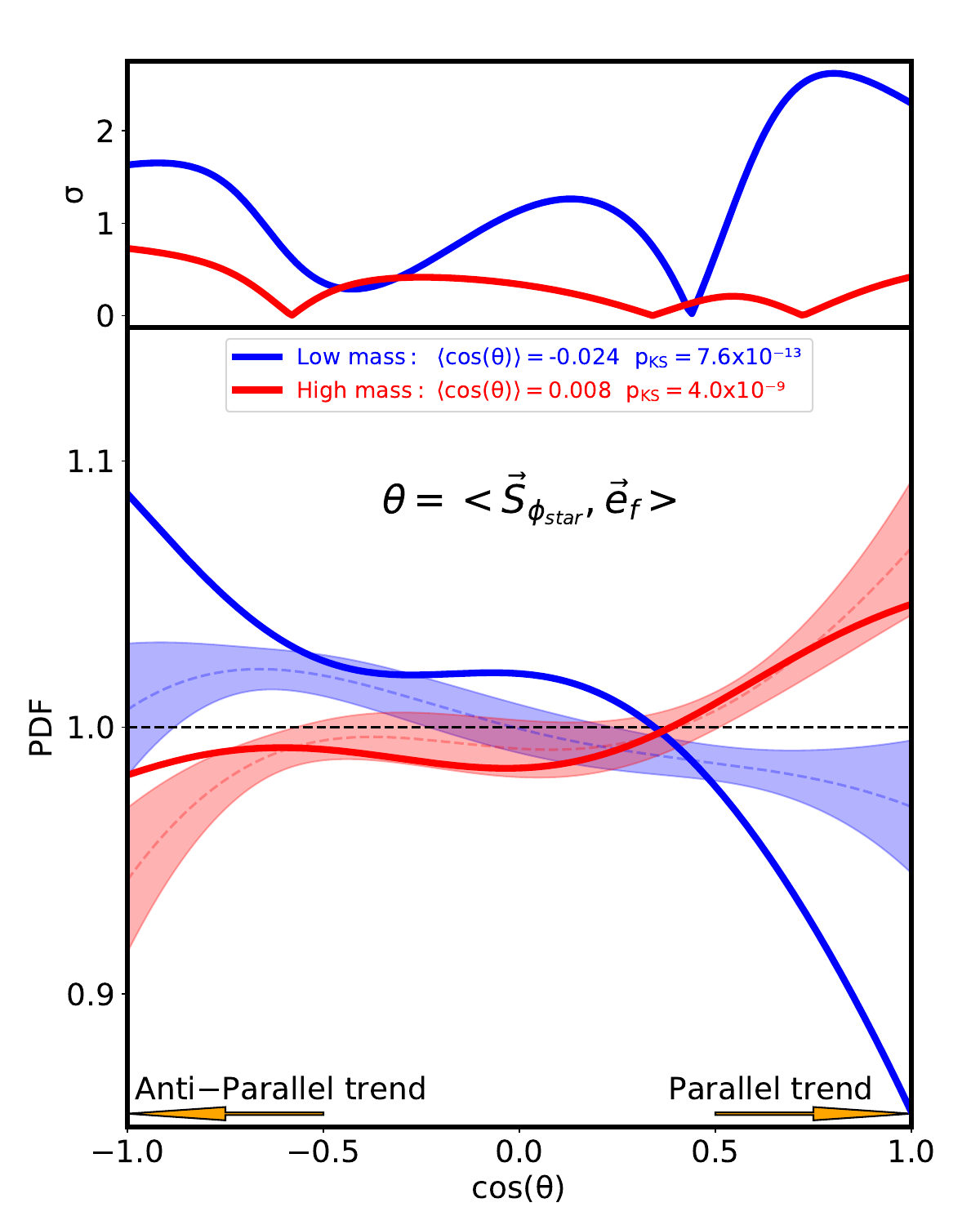}{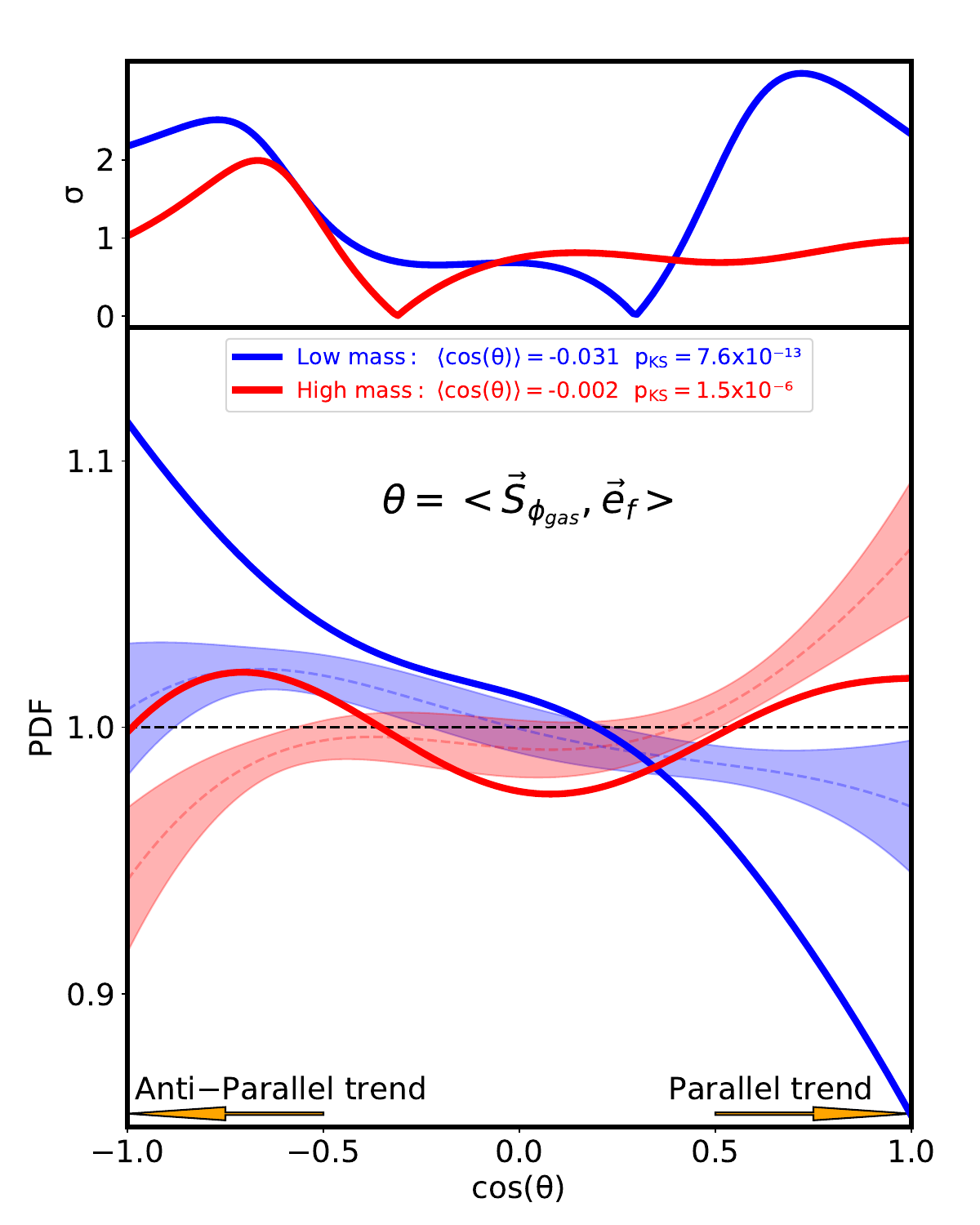}
    \caption{The \textbf{bottom subfigures} in each panel show the probability distribution of $\rm cos(\theta)$, where $\theta$ is the angle between the galaxy spin and the filament spin. The \textbf{left panel} presents the spin of the stellar component of galaxies, while the \textbf{right panel} shows the spin of the gas component, as indicated at the top center of each panel. Blue and red solid lines represent low-mass and high-mass galaxies, respectively. The horizontal black dashed line denotes the expected distribution for a uniform random orientation of $\rm cos(\theta)$. Color-filled regions indicate the $1\sigma$ statistical uncertainties derived from 10,000 null-hypothesis realizations. The mean value of $\rm cos(\theta)$ and the p-value from the Kolmogorov–Smirnov (KS) test ($\rm p_{KS}$) are displayed at the top center. Aligned and anti-aligned trends are highlighted with yellow arrows. The \textbf{upper subfigures} in each panel show the discrepancy, in standard deviation units of the randomized distribution, between the observed curve and the mean of the random distributions as a function of $\rm cos(\theta)$.}
    \label{fig:f3}
\end{figure*}

Before analyzing the correlation between galaxy spin and filament spin directions, we first aim to reproduce previous studies that examine the absolute cosine of the alignment angle between these vectors. As shown in Figure~\ref{fig:f2}, we evaluate the absolute cosine values ($\rm |\cos(\theta)|$) of the angle between galaxy spin and the spine of their associated large-scale filaments as a function of stellar mass.

The blue curve with star symbols in Figure~\ref{fig:f2} illustrates the alignment correlation, $\rm |\cos(\theta)|$, between the spin of the stellar component of galaxies and the spine of their host filament, revealing a mass-dependent transition. This correlation systematically weakens with increasing stellar mass. As indicated by the upward yellow arrow, low-mass galaxies exhibit a statistically significant alignment, with their spins oriented parallel to the filament. In contrast, high-mass galaxies transition to an perpendicular configuration, with spins perpendicular to the filament, as denoted by the downward yellow arrow. The critical transition occurs near $\rm log_{10} M_\star/M_\odot \sim 10 $. This overall trend is consistent with findings from previous studies in observations\citep{2020MNRAS.491.2864W} and simulations \citep{2007MNRAS.381...41H,2007ApJ...655L...5A,2012MNRAS.427.3320C,2015MNRAS.452.3369C, 2014MNRAS.444.1453D,2015MNRAS.452.3369C, 2018MNRAS.481..414G,2019MNRAS.483.3227K,2019MNRAS.487.1607G,2020OJAp....3E...3N,2021MNRAS.503.2280G,2021MNRAS.502.5528L}.

We also examined the correlation between the spin of the gas component and the filament structure. As shown by the red solid line with dot symbols in Figure~\ref{fig:f2}, within the margin of error, no significant difference is observed between the measured signal and the random signal, represented by the dotted black line. This suggests that, based on the current sample, the spin of the gas component exhibits little to no correlation with the filaments.

We now directly examine the correlation between galaxy spin vectors and filament spin directions, quantified by $\rm \cos(\theta)$. Figure~\ref{fig:f3} presents results for both low-mass (blue solid lines) and high-mass (red solid lines) galaxies, considering the spins of their stellar components (left panel) and gas components (right panel).

\begin{figure*}
    \centering
    \plottwo{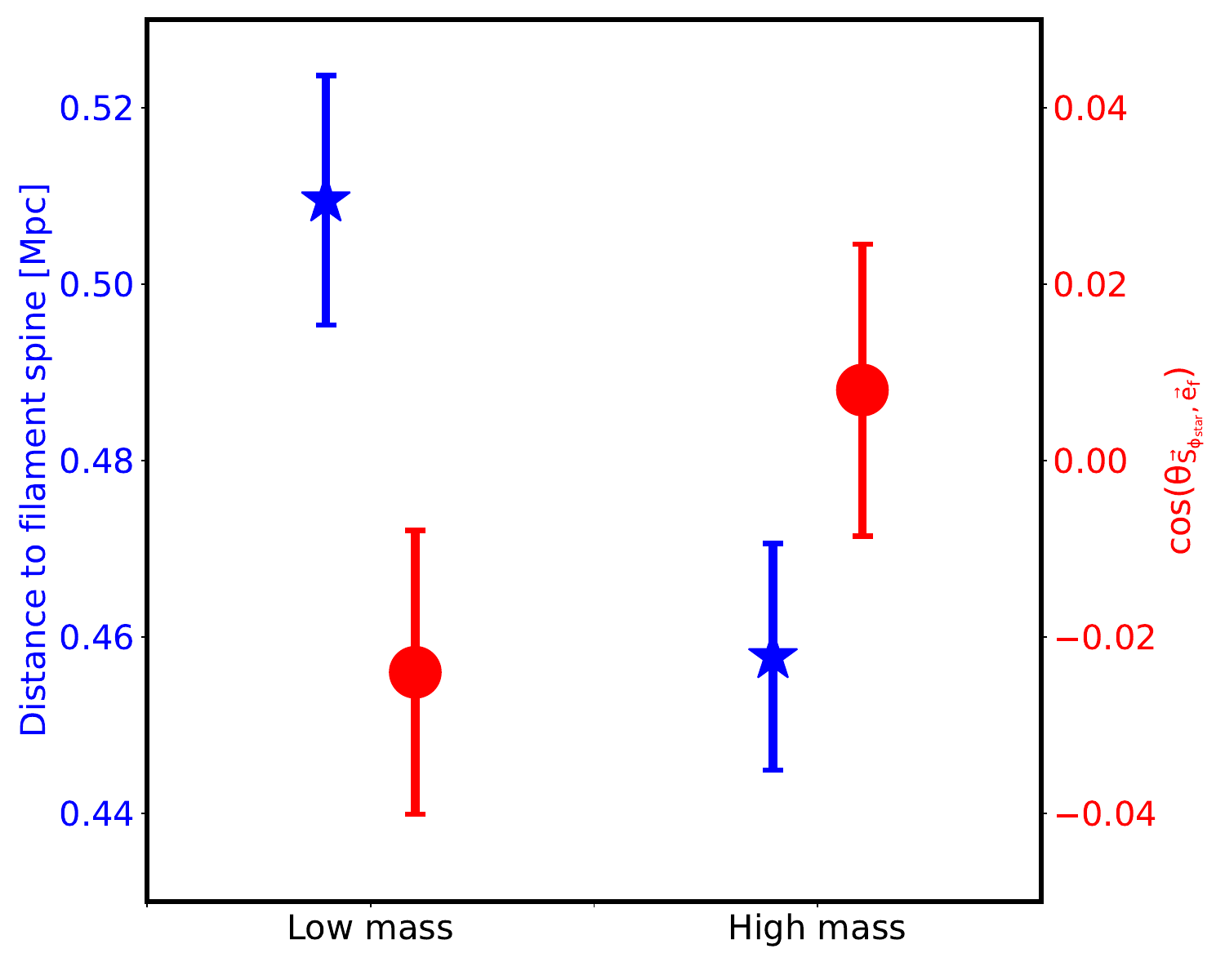}{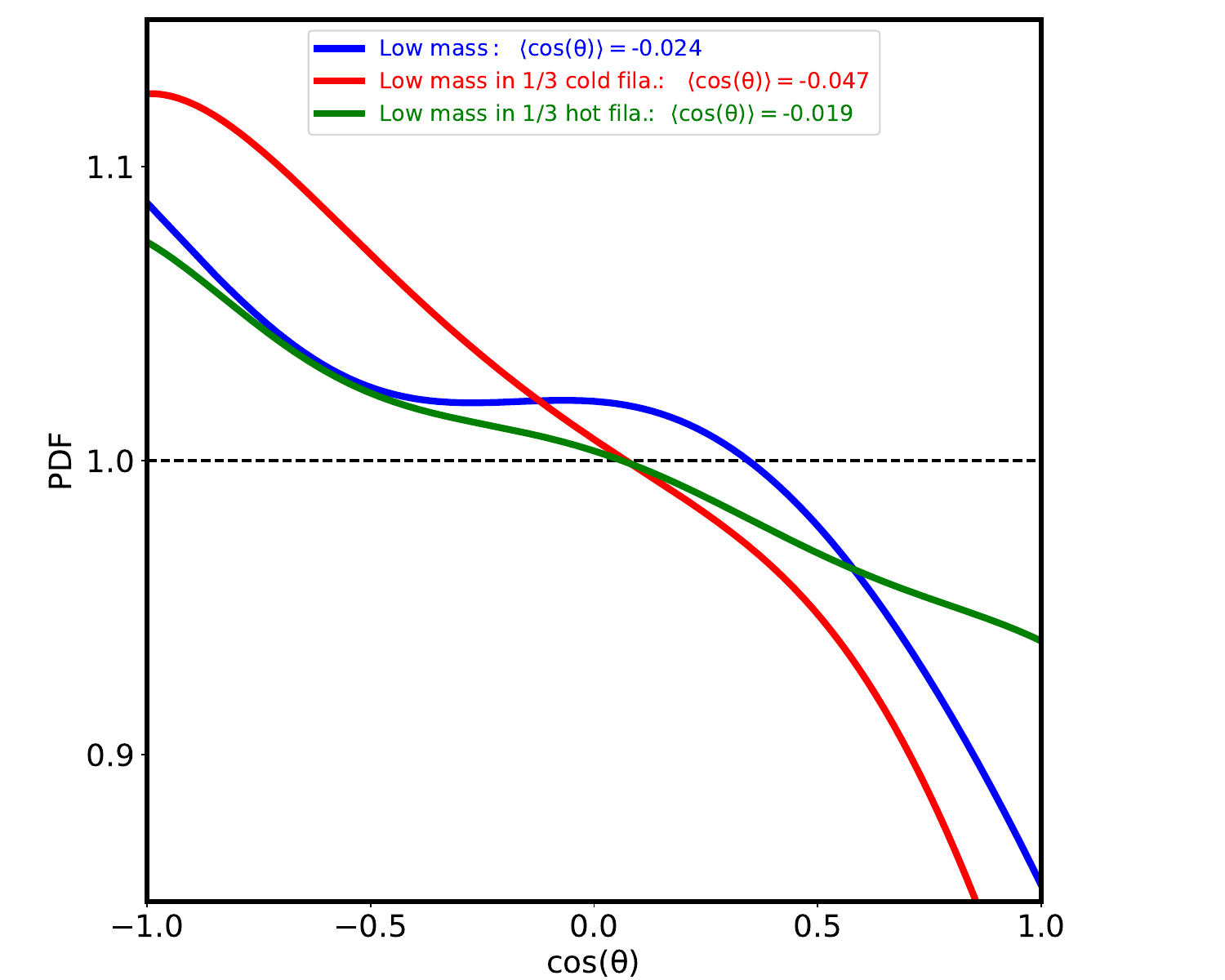}
    \caption{\textbf{Left panel:} the average distance (represented by blue lines with star symbols, corresponding to the left y-axis) and  $\rm cos(\theta)$values (shown in red lines with solid circles, corresponding to the right y-axis) are plotted for both low-mass and high-mass galaxies. Error bars indicate the 1$\sigma$ uncertainties.
    \textbf{Right panel}: similar to the bottom-left panel fo Fig.~\ref{fig:f3}, but show the effect of filament dynamical temperature. The low-mass galaxy sample is divided into two subsamples based on the dynamical temperature of their host filaments. The red and green lines represent the lowest and highest one-third of the sample in terms of filament dynamical temperature, respectively.
    }
    \label{fig:f4}
\end{figure*}

To evaluate the statistical significance of the observed correlation, we follow the method used in \cite{2013ApJ...775L..42T, 2015ApJ...798...17Z, 2016MNRAS.457..695P} and \cite{2025ApJ...982L..29B}. We generate 10,000 random galaxy samples. Each random sample contains the same number of galaxies as the observed MaNGA sample and preserves the original filament spin directions. In each random trial, the galaxies positions (namely RA, Dec, z, b/a) are fixed, but their position angles (either  $\phi_{gas}$ or $\phi_{star}$) are randomly assigned between $0^{\circ}$ and $360^{\circ}$. Following the procedures described above, we recalculate the mean value of $\cos(\theta_R)$ for each mock sample, resulting in a distribution of 10,000 mean values of $\cos(\theta_R)$. The standard deviation of this distribution is then used to quantify the significance of the observed signal via 
\[
\mathrm{significance} = \frac{|\cos(\theta_{obs}) - \cos(\theta_R)|}{\sigma(cos(\theta_R))}
\]

Figure~\ref{fig:f3} reveals two salient trends: first, low-mass galaxies exhibit a statistically significant tendency toward anti-parallel alignment with filaments, with approaching $3\sigma$ respect to the random distribution at some values of $\rm \cos(\theta)$, whereas high-mass galaxies show only a modest trend toward parallel alignment within $1\sigma$, approaching randomness. This trend is not against what we conclude in Fig.\ref{fig:f2} where massive galaxies tend to be perpendicular, considering that the mass threshold we adopt here is lower than the transition mass in Fig.\ref{fig:f2}, the mixture of parallel and perpendicular sub-samples will yield a random 
result. Second, a comparison between the panels indicates that the gas component (blue line in the right panel) displays a slightly stronger anti-parallel signal.

Previous studies \citep{2018MNRAS.474..547K, 2018MNRAS.474.5437L, 2024MNRAS.528.4139H} have suggested that galaxy properties correlate with their distance from the filament spine. 
As illustrated in Figure~\ref{fig:f4}, we analyze both the distances of two galaxy subsamples from the filament spine (left panel) and the impact of filament dynamical temperature on the spin-filament alignment of low-mass galaxies (right panel). The results show that high-mass spiral galaxies tend to reside closer to the filament spine, whereas low-mass spiral galaxies are generally found farther away. Furthermore, the probability distribution function of $\cos(\theta)$ reveals that low-mass galaxies in cold filaments exhibit a stronger anti-alignment signal ($\langle \cos(\theta) \rangle = -0.047$ shown in red line) compared to those in hot filaments ($\langle \cos(\theta) \rangle = -0.019$ shown in blue line), indicating that the dynamical state of filaments plays a significant role in modulating the spin-filament alignment. Taken together, these findings suggest that the spin directions of spiral galaxies located outside the filaments, especially those in colder filaments, tend to be more anti-parallel to the filament spin directions---a point that will be elaborated upon in the discussion.

\section{Summary and Discussion}\label{sec:s&d}

In this study, we investigated the alignment between the spin of stellar/gas components and their hosting filament using a sample of spiral galaxies selected from the MaNGA survey. Our key findings are summarized as follows:

\begin{enumerate}
    \item When examining the absolute cosine values (quantified by $|\cos(\theta)|$) representing spin alignment relative to filament axes, stellar components exhibit a mass-dependent correlation with filament orientation. Low-mass spiral galaxies preferentially show parallel alignment, while high-mass counterparts demonstrate a transition to perpendicular trend. In contrast, gas components display no systematic correlation with filament axes, consistent with random orientations across the current sample.
    \item Upon incorporating filament spin properties (quantified by $\cos(\theta)$), low-mass spirals exhibit a statistically significant anti-parallel trend between spins and filament spins. This anti-parallel is particularly pronounced in the gas component compared to stars. No such anti-parallel signature is detected in high-mass spiral galaxies.
    \item Spatial distribution analysis reveals systematic environmental dependence — high-mass spiral galaxies preferentially reside closer to filament spines, whereas their low-mass counterparts tend to occupy regions at greater distances from the central filamentary structures.
\end{enumerate}

For the first time, we have detected a transition in galaxy spin-filament correlation from parallel to perpendicular of  spiral galaxies. While our analysis of absolute values, $|\cos(\theta)|$, between spiral galaxy spin and filament are broadly supports conclusions from earlier studies \citep{2020MNRAS.491.2864W, 2022MNRAS.516.3569B}. The current sample shows no statistically significant signal in the spin of gas component. This observational outcome exhibits partial consistency with results from SAMI \citep{2020MNRAS.491.2864W}, which reports a transition mass between $10^{10.4}$ and $\rm 10^{10.9} M_\odot$, Illustris-1 \citep{2018ApJ...866..138W} which claims a transition mass around $\rm \sim10^{9.4}h^{-1}M_\odot$ and Horizon-AGN simulation \citep{2018MNRAS.481.4753C} which suggests that the transition occurs at $\rm \sim10^{10.1\pm0.3}M_\odot$,  reporting comparable transition mass scales ($\rm \sim10^{10}M_\odot$), though direct morphological comparisons were absent in prior work. Divergences in alignment strength between studies may reflect variations in sample selection methodologies. The weaker signals we observed compared to hydrodynamic simulations may be due to factors like measurement errors and other inherent challenges in determining galaxy spin from observations.

When considering the correlation between the two vectors, that is, the parallel and anti-parallel correlation between the directions of galaxy spin and the filament spin, we found that low-mass spiral galaxies and high-mass spiral galaxies exhibit inconsistent behavior. The spin of low-mass spiral galaxies is anti-parallel to the filament spin direction. This discovery forces us to think more deeply about how spiral galaxies obtain spin from their surroundings. We further find that the anti-parallel alignment of low-mass spiral galaxies is significantly stronger in colder filaments. This indicates that the dynamical state of filaments can modulate the spin-filament correlation, and highlights the importance of considering filament properties when interpreting these alignments.

Based on our current sample, we find that high-mass spiral galaxies are preferentially located closer to the centers of filaments, while low-mass spiral galaxies are more likely to reside in the outskirts. 
This finding agree with previous conclusion \citep{2018MNRAS.474.5437L, 2018MNRAS.474..547K, 2024MNRAS.528.4139H, zhang2025galaxyhalopropertiescosmic} that the closer to the center of the filament, the greater the mass of the galaxy. 
This spatial segregation of spiral galaxies relative to filaments may offer useful clues about their spin alignment correlation.
High-mass spiral galaxies located near filament cores are likely embedded in environments dominated by strong large-scale tidal fields, where material flows predominantly along the filament spine. Local interactions, such as group accretion or minor mergers, become more significant, thereby diluting or modifying spin alignment signals.  In contrast, low-mass spiral galaxies residing in the outskirts experience weaker tidal coherence; in these regions, material flows may be more isotropic or preferentially directed perpendicular to the filament spine. As a result, low-mass spiral galaxies are less perturbed by local dynamical interactions and can better preserve the primordial spin alignments imparted by large-scale tidal torques, leading to a clearer correlation between their spin directions and the filament orientation. However, we note that the tidal field in the cosmic web is highly complex and cannot be fully captured by a simple picture of tidal coherence between filament spines and their outskirts. Recent studies, such as \citep{2024arXiv240716489K}, have demonstrated that the tidal influence of voids and other large-scale structures can play a significant role even within filamentary regions. Therefore, the observed alignment trends may result from a combination of multiple environmental factors, and a more comprehensive understanding of the underlying tidal configurations will require further dedicated investigation.

Previous studies have proposed diverse mechanisms to explain the origin of galaxy spin-filament alignments. The tidal torque theory \citep{Hoyle1949,1969ApJ...155..393P,1970Ap......6..320D,1979MNRAS.186..133E,1984ApJ...286...38W, 1987ApJ...319..575B,2007JRSSC..56....1S} attributes galaxy spin to early tidal interactions, while later work identifies non-universal mass accretion patterns as critical drivers. Detailed studies about the TTT \citep{2000ApJ...532L...5L,2002MNRAS.332..325P,2002MNRAS.332..339P,2018MNRAS.481..414G,2021MNRAS.503.2280G, 2021MNRAS.502.5528L,2024PASP..136c7001L,2025arXiv250501298L} investigated different aspects of the effects of tidal field on galaxies, related to the establishment and alignment of galactic angular momenta.
\citet{2014MNRAS.443.1274L, 2015ApJ...807...37S,2015ApJ...813....6K} suggest universal filament-aligned accretion for massive halos, this fails for low-mass systems. 
\citet{2012MNRAS.427.3320C,2014MNRAS.445L..46W,2015MNRAS.446.2744L} pointed out that low-mass haloes are mainly formed by
smoothing accretion through the wind of flows embedded in mis-aligned walls, and massive haloes are products of major mergers in
filaments. 
\cite{2018MNRAS.473.1562W} resolve this dichotomy via a two-stage model: the spin of the dark halo was initially parallel to the slowest collapsed direction, and the change in the spin of the dark halo came from the change in the way the dark halo accreted matter from the surrounding environment. This explains the mass-dependent flip in spin-filament correlations. Recent studies \citep{2020PhRvL.124j1302Y, 2023ApJ...943..128S} have proposed investigating the angular momentum of galaxies by tracing their origins back to the very early universe.

However, they did not distinguish between positive parallel and anti-parallel correlations. Our results represent the first observational discovery of anti-parallel correlation signals in low-mass spiral galaxies. This finding enables a re-examination of the mechanisms behind galaxy spin formation and the influence of filament spin on galaxy spin, that connect accretion physics, environmental transitions, and multi-scale cosmic flows. 

This discovery also prompts us to consider the classic ``Nature or Nurture'' question in the context of galaxy spin. One plausible explanation for the observed anti-parallel correlations is that they may originate from intrinsic properties of the cosmic web, where filaments and galaxies inherit anti-aligned angular momenta during large-scale structure formation, which can be seen as a ``nature'' - like factor. Alternatively, the anti-parallel alignment could emerge through the acquired effect of the cumulative anti-aligned orbital angular momentum of accreted material over time, representing a ``nurture'' - like process.

One feasible approach is to select two sets of samples from the simulation at $z=0$: one parallel and one anti-parallel, and trace back their formation histories to identify differences in their origin and evolution. Similarly, we can select two sets of samples at higher redshifts to investigate their subsequent evolutionary paths. These investigations will be left for future work.

\begin{acknowledgments}
%The authors thank anonymous referees for comments that substantially improved the manuscript.  
The authors thank Noam Libeskind and Elmo Tempel for their useful discussions.
PW, WHD and XXT acknowledge the financial support from the NSFC (No.12473009), and also sponsored by Shanghai Rising-Star Program (No.24QA2711100). YCZ acknowledges the financial support from the NSFC (No.12273088). M.B. acknowledges support by the National Natural Science Foundation of China, NSFC grant No. 12303009.

\end{acknowledgments}

\bibliography{main}{}

\begin{thebibliography}{}
\expandafter\ifx\csname natexlab\endcsname\relax\def\natexlab#1{#1}\fi
\providecommand{\url}[1]{\href{#1}{#1}}
\providecommand{\dodoi}[1]{doi:~\href{http://doi.org/#1}{\nolinkurl{#1}}}
\providecommand{\doeprint}[1]{\href{http://ascl.net/#1}{\nolinkurl{http://ascl.net/#1}}}
\providecommand{\doarXiv}[1]{\href{https://arxiv.org/abs/#1}{\nolinkurl{https://arxiv.org/abs/#1}}}

\bibitem[{Alam {et~al.}(2015)Alam, Albareti, Allende~Prieto, Anders, Anderson, Anderton, Andrews, Armengaud, Aubourg, Bailey, Basu, Bautista, Beaton, Beers, Bender, Berlind, Beutler, Bhardwaj, Bird, Bizyaev, Blake, Blanton, Blomqvist, Bochanski, Bolton, Bovy, Shelden~Bradley, Brandt, Brauer, Brinkmann, Brown, Brownstein, Burden, Burtin, Busca, Cai, Capozzi, Carnero~Rosell, Carr, Carrera, Chambers, Chaplin, Chen, Chiappini, Chojnowski, Chuang, Clerc, Comparat, Covey, Croft, Cuesta, Cunha, da~Costa, Da~Rio, Davenport, Dawson, De~Lee, Delubac, Deshpande, Dhital, Dutra-Ferreira, Dwelly, Ealet, Ebelke, Edmondson, Eisenstein, Ellsworth, Elsworth, Epstein, Eracleous, Escoffier, Esposito, Evans, Fan, Fernández-Alvar, Feuillet, Filiz~Ak, Finley, Finoguenov, Flaherty, Fleming, Font-Ribera, Foster, Frinchaboy, Galbraith-Frew, García, García-Hernández, García~Pérez, Gaulme, Ge, Génova-Santos, Georgakakis, Ghezzi, Gillespie, Girardi, Goddard, Gontcho, González~Hernández, Grebel, Green, {et~al.}}]{SDSS-III}
Alam, S., Albareti, F.~D., Allende~Prieto, C., {et~al.} 2015, The Astrophysical Journal Supplement Series, 219, 12, \dodoi{10.1088/0067-0049/219/1/12}

\bibitem[{{Arag{\'o}n-Calvo} {et~al.}(2007){Arag{\'o}n-Calvo}, {van de Weygaert}, {Jones}, \& {van der Hulst}}]{2007ApJ...655L...5A}
{Arag{\'o}n-Calvo}, M.~A., {van de Weygaert}, R., {Jones}, B. J.~T., \& {van der Hulst}, J.~M. 2007, \apjl, 655, L5, \dodoi{10.1086/511633}

\bibitem[{{Bao} {et~al.}(2025){Bao}, {Chen}, {Gu}, {Wang}, {Shi}, \& {Wang}}]{2025ApJ...982L..29B}
{Bao}, M., {Chen}, Y., {Gu}, Q., {et~al.} 2025, \apjl, 982, L29, \dodoi{10.3847/2041-8213/adbc68}

\bibitem[{Barnes \& Efstathiou(1987)}]{1987ApJ...319..575B}
Barnes, J., \& Efstathiou, G. 1987, The Astrophysical Journal, 319, 575, \dodoi{10.1086/165480}

\bibitem[{{Barsanti} {et~al.}(2022){Barsanti}, {Colless}, {Welker}, {Oh}, {Casura}, {Bryant}, {Croom}, {D'Eugenio}, {Lawrence}, {Richards}, \& {van de Sande}}]{2022MNRAS.516.3569B}
{Barsanti}, S., {Colless}, M., {Welker}, C., {et~al.} 2022, \mnras, 516, 3569, \dodoi{10.1093/mnras/stac2405}

\bibitem[{{Barsanti} {et~al.}(2023){Barsanti}, {Colless}, {D'Eugenio}, {Oh}, {Bryant}, {Casura}, {Croom}, {Mai}, {Ristea}, {van de Sande}, {Welker}, \& {Zovaro}}]{2023MNRAS.526.1613B}
{Barsanti}, S., {Colless}, M., {D'Eugenio}, F., {et~al.} 2023, \mnras, 526, 1613, \dodoi{10.1093/mnras/stad2728}

\bibitem[{{Barsanti} {et~al.}(2025){Barsanti}, {Croom}, {Colless}, {Bland-Hawthorn}, {Brough}, {Bryant}, {Lorente}, {Oh}, {Santucci}, {Sweet}, {Sande van de}, \& {Welker}}]{2025MNRAS.538.2660B}
{Barsanti}, S., {Croom}, S.~M., {Colless}, M., {et~al.} 2025, \mnras, 538, 2660, \dodoi{10.1093/mnras/staf426}

\bibitem[{Bundy {et~al.}(2015)Bundy, Bershady, Law, Yan, Drory, MacDonald, Wake, Cherinka, Sánchez-Gallego, Weijmans, Thomas, Tremonti, Masters, Coccato, Diamond-Stanic, Aragón-Salamanca, Avila-Reese, Badenes, Falcón-Barroso, Belfiore, Bizyaev, Blanc, Bland-Hawthorn, Blanton, Brownstein, Byler, Cappellari, Conroy, Dutton, Emsellem, Etherington, Frinchaboy, Fu, Gunn, Harding, Johnston, Kauffmann, Kinemuchi, Klaene, Knapen, Leauthaud, Li, Lin, Maiolino, Malanushenko, Malanushenko, Mao, Maraston, McDermid, Merrifield, Nichol, Oravetz, Pan, Parejko, Sanchez, Schlegel, Simmons, Steele, Steinmetz, Thanjavur, Thompson, Tinker, van~den Bosch, Westfall, Wilkinson, Wright, Xiao, \& Zhang}]{2015ApJ...798....7B}
Bundy, K., Bershady, M.~A., Law, D.~R., {et~al.} 2015, The Astrophysical Journal, 798, 7, \dodoi{10.1088/0004-637x/798/1/7}

\bibitem[{{Cautun} {et~al.}(2014){Cautun}, {van de Weygaert}, {Jones}, \& {Frenk}}]{2014MNRAS.441.2923C}
{Cautun}, M., {van de Weygaert}, R., {Jones}, B. J.~T., \& {Frenk}, C.~S. 2014, \mnras, 441, 2923, \dodoi{10.1093/mnras/stu768}

\bibitem[{{Chen} {et~al.}(2016){Chen}, {Shi}, {Tremonti}, {Bershady}, {Merrifield}, {Emsellem}, {Jin}, {Huang}, {Fu}, {Wake}, {Bundy}, {Stark}, {Lin}, {Argudo-Fernandez}, {Bergmann}, {Bizyaev}, {Brownstein}, {Bureau}, {Chisholm}, {Drory}, {Guo}, {Hao}, {Hu}, {Li}, {Li}, {Roman Lopes}, {Pan}, {Riffel}, {Thomas}, {Wang}, {Westfall}, \& {Yan}}]{2016NatCo...713269C}
{Chen}, Y.-M., {Shi}, Y., {Tremonti}, C.~A., {et~al.} 2016, Nature Communications, 7, 13269, \dodoi{10.1038/ncomms13269}

\bibitem[{{Codis}(2016)}]{2016arXiv161206873C}
{Codis}, S. 2016, arXiv e-prints, arXiv:1612.06873, \dodoi{10.48550/arXiv.1612.06873}

\bibitem[{{Codis} {et~al.}(2018){Codis}, {Jindal}, {Chisari}, {Vibert}, {Dubois}, {Pichon}, \& {Devriendt}}]{2018MNRAS.481.4753C}
{Codis}, S., {Jindal}, A., {Chisari}, N.~E., {et~al.} 2018, \mnras, 481, 4753, \dodoi{10.1093/mnras/sty2567}

\bibitem[{{Codis} {et~al.}(2012){Codis}, {Pichon}, {Devriendt}, {Slyz}, {Pogosyan}, {Dubois}, \& {Sousbie}}]{2012MNRAS.427.3320C}
{Codis}, S., {Pichon}, C., {Devriendt}, J., {et~al.} 2012, \mnras, 427, 3320, \dodoi{10.1111/j.1365-2966.2012.21636.x}

\bibitem[{{Codis} {et~al.}(2015){Codis}, {Pichon}, \& {Pogosyan}}]{2015MNRAS.452.3369C}
{Codis}, S., {Pichon}, C., \& {Pogosyan}, D. 2015, \mnras, 452, 3369, \dodoi{10.1093/mnras/stv1570}

\bibitem[{Croom {et~al.}(2021)Croom, Owers, Scott, Poetrodjojo, Groves, van~de Sande, Barone, Cortese, D'Eugenio, Bland-Hawthorn, Bryant, Oh, Brough, Agostino, Casura, Catinella, Colless, Cecil, Davies, Drinkwater, Driver, Ferreras, Foster, Fraser-McKelvie, Lawrence, Leslie, Liske, López-Sánchez, Lorente, McElroy, Medling, Obreschkow, Richards, Sharp, Sweet, Taranu, Taylor, Tescari, Thomas, Tocknell, \& Vaughan}]{2021MNRAS.505..991C}
Croom, S.~M., Owers, M.~S., Scott, N., {et~al.} 2021, Monthly Notices of the Royal Astronomical Society, 505, 991, \dodoi{10.1093/mnras/stab229}

\bibitem[{{Doroshkevich}(1970)}]{1970Ap......6..320D}
{Doroshkevich}, A.~G. 1970, Astrophysics, 6, 320, \dodoi{10.1007/BF01001625}

\bibitem[{{Dubois} {et~al.}(2014){Dubois}, {Pichon}, {Welker}, {Le Borgne}, {Devriendt}, {Laigle}, {Codis}, {Pogosyan}, {Arnouts}, {Benabed}, {Bertin}, {Blaizot}, {Bouchet}, {Cardoso}, {Colombi}, {de Lapparent}, {Desjacques}, {Gavazzi}, {Kassin}, {Kimm}, {McCracken}, {Milliard}, {Peirani}, {Prunet}, {Rouberol}, {Silk}, {Slyz}, {Sousbie}, {Teyssier}, {Tresse}, {Treyer}, {Vibert}, \& {Volonteri}}]{2014MNRAS.444.1453D}
{Dubois}, Y., {Pichon}, C., {Welker}, C., {et~al.} 2014, \mnras, 444, 1453, \dodoi{10.1093/mnras/stu1227}

\bibitem[{{Efstathiou} \& {Jones}(1979)}]{1979MNRAS.186..133E}
{Efstathiou}, G., \& {Jones}, B.~J.~T. 1979, \mnras, 186, 133, \dodoi{10.1093/mnras/186.2.133}

\bibitem[{Eisenstein {et~al.}(2011)Eisenstein, Weinberg, Agol, Aihara, Allende~Prieto, Anderson, Arns, Aubourg, Bailey, Balbinot, Barkhouser, Beers, Berlind, Bickerton, Bizyaev, Blanton, Bochanski, Bolton, Bosman, Bovy, Brandt, Breslauer, Brewington, Brinkmann, Brown, Brownstein, Burger, Busca, Campbell, Cargile, Carithers, Carlberg, Carr, Chang, Chen, Chiappini, Comparat, Connolly, Cortes, Croft, Cunha, da~Costa, Davenport, Dawson, De~Lee, Porto~de Mello, de~Simoni, Dean, Dhital, Ealet, Ebelke, Edmondson, Eiting, Escoffier, Esposito, Evans, Fan, Femenía~Castellá, Dutra~Ferreira, Fitzgerald, Fleming, Font-Ribera, Ford, Frinchaboy, García~Pérez, Gaudi, Ge, Ghezzi, Gillespie, Gilmore, Girardi, Gott, Gould, Grebel, Gunn, Hamilton, Harding, Harris, Hawley, Hearty, Hennawi, González~Hernández, Ho, Hogg, Holtzman, Honscheid, Inada, Ivans, Jiang, Jiang, Johnson, Jordan, Jordan, Kauffmann, Kazin, Kirkby, Klaene, Knapp, Kneib, Kochanek, {et~al.}}]{Eisenstein2011}
Eisenstein, D.~J., Weinberg, D.~H., Agol, E., {et~al.} 2011, The Astronomical Journal, 142, 72, \dodoi{10.1088/0004-6256/142/3/72}

\bibitem[{{Ganeshaiah Veena} {et~al.}(2019){Ganeshaiah Veena}, {Cautun}, {Tempel}, {van de Weygaert}, \& {Frenk}}]{2019MNRAS.487.1607G}
{Ganeshaiah Veena}, P., {Cautun}, M., {Tempel}, E., {van de Weygaert}, R., \& {Frenk}, C.~S. 2019, \mnras, 487, 1607, \dodoi{10.1093/mnras/stz1343}

\bibitem[{{Ganeshaiah Veena} {et~al.}(2021){Ganeshaiah Veena}, {Cautun}, {van de Weygaert}, {Tempel}, \& {Frenk}}]{2021MNRAS.503.2280G}
{Ganeshaiah Veena}, P., {Cautun}, M., {van de Weygaert}, R., {Tempel}, E., \& {Frenk}, C.~S. 2021, \mnras, 503, 2280, \dodoi{10.1093/mnras/stab411}

\bibitem[{{Ganeshaiah Veena} {et~al.}(2018){Ganeshaiah Veena}, {Cautun}, {van de Weygaert}, {Tempel}, {Jones}, {Rieder}, \& {Frenk}}]{2018MNRAS.481..414G}
{Ganeshaiah Veena}, P., {Cautun}, M., {van de Weygaert}, R., {et~al.} 2018, \mnras, 481, 414, \dodoi{10.1093/mnras/sty2270}

\bibitem[{Gunn {et~al.}(2006)Gunn, Siegmund, Mannery, Owen, Hull, Leger, Carey, Knapp, York, Boroski, Kent, Lupton, Rockosi, Evans, Waddell, Anderson, Annis, Barentine, Bartoszek, Bastian, Bracker, Brewington, Briegel, Brinkmann, Brown, Carr, Czarapata, Drennan, Dombeck, Federwitz, Gillespie, Gonzales, Hansen, Harvanek, Hayes, Jordan, Kinney, Klaene, Kleinman, Kron, Kresinski, Lee, Limmongkol, Lindenmeyer, Long, Loomis, McGehee, Mantsch, Neilsen, Neswold, Newman, Nitta, Peoples, Pier, Prieto, Prosapio, Rivetta, Schneider, Snedden, \& Wang}]{2006AJ....131.2332G}
Gunn, J.~E., Siegmund, W.~A., Mannery, E.~J., {et~al.} 2006, The Astronomical Journal, 131, 2332, \dodoi{10.1086/500975}

\bibitem[{{Hahn} {et~al.}(2007){Hahn}, {Carollo}, {Porciani}, \& {Dekel}}]{2007MNRAS.381...41H}
{Hahn}, O., {Carollo}, C.~M., {Porciani}, C., \& {Dekel}, A. 2007, \mnras, 381, 41, \dodoi{10.1111/j.1365-2966.2007.12249.x}

\bibitem[{Hahn {et~al.}(2007)Hahn, Porciani, Carollo, \& Dekel}]{2007MNRAS.375..489H}
Hahn, O., Porciani, C., Carollo, C.~M., \& Dekel, A. 2007, Monthly Notices of the Royal Astronomical Society, 375, 489, \dodoi{10.1111/j.1365-2966.2006.11318.x}

\bibitem[{{Hahn} {et~al.}(2009){Hahn}, {Porciani}, {Dekel}, \& {Carollo}}]{2009MNRAS.398.1742H}
{Hahn}, O., {Porciani}, C., {Dekel}, A., \& {Carollo}, C.~M. 2009, \mnras, 398, 1742, \dodoi{10.1111/j.1365-2966.2009.15271.x}

\bibitem[{{Hoffman} {et~al.}(2012){Hoffman}, {Metuki}, {Yepes}, {Gottl{\"o}ber}, {Forero-Romero}, {Libeskind}, \& {Knebe}}]{2012MNRAS.425.2049H}
{Hoffman}, Y., {Metuki}, O., {Yepes}, G., {et~al.} 2012, \mnras, 425, 2049, \dodoi{10.1111/j.1365-2966.2012.21553.x}

\bibitem[{{Hoosain} {et~al.}(2024){Hoosain}, {Blyth}, {Skelton}, {Kannappan}, {Stark}, {Eckert}, {Hutchens}, {Carr}, \& {Kraljic}}]{2024MNRAS.528.4139H}
{Hoosain}, M., {Blyth}, S.-L., {Skelton}, R.~E., {et~al.} 2024, \mnras, 528, 4139, \dodoi{10.1093/mnras/stae174}

\bibitem[{{Hoyle} {et~al.}(1949){Hoyle}, {Burgers}, \& {van de Hulst}}]{Hoyle1949}
{Hoyle}, F., {Burgers}, J., M., \& {van de Hulst}, H., C. 1949, eds., in Problems of Cosmical Aerodynamics, Central Air Documents Office, Dayton, p. 195. coso

\bibitem[{{Huchra} {et~al.}(2012){Huchra}, {Macri}, {Masters}, {Jarrett}, {Berlind}, {Calkins}, {Crook}, {Cutri}, {Erdo{\v{g}}du}, {Falco}, {George}, {Hutcheson}, {Lahav}, {Mader}, {Mink}, {Martimbeau}, {Schneider}, {Skrutskie}, {Tokarz}, \& {Westover}}]{2012ApJS..199...26H}
{Huchra}, J.~P., {Macri}, L.~M., {Masters}, K.~L., {et~al.} 2012, \apjs, 199, 26, \dodoi{10.1088/0067-0049/199/2/26}

\bibitem[{{Jin} {et~al.}(2016){Jin}, {Chen}, {Shi}, {Tremonti}, {Bershady}, {Merrifield}, {Emsellem}, {Fu}, {Wake}, {Bundy}, {Lin}, {Argudo-Fernandez}, {Huang}, {Stark}, {Storchi-Bergmann}, {Bizyaev}, {Brownstein}, {Chisholm}, {Guo}, {Hao}, {Hu}, {Li}, {Li}, {Masters}, {Malanushenko}, {Pan}, {Riffel}, {Roman-Lopes}, {Simmons}, {Thomas}, {Wang}, {Westfall}, \& {Yan}}]{2016MNRAS.463..913J}
{Jin}, Y., {Chen}, Y., {Shi}, Y., {et~al.} 2016, \mnras, 463, 913, \dodoi{10.1093/mnras/stw2055}

\bibitem[{{Kang} \& {Wang}(2015)}]{2015ApJ...813....6K}
{Kang}, X., \& {Wang}, P. 2015, \apj, 813, 6, \dodoi{10.1088/0004-637X/813/1/6}

\bibitem[{Kraljic {et~al.}(2021)Kraljic, Duckworth, Tojeiro, Alam, Bizyaev, Weijmans, Boardman, \& Lane}]{2021MNRAS.504.4626K}
Kraljic, K., Duckworth, C., Tojeiro, R., {et~al.} 2021, Monthly Notices of the Royal Astronomical Society, 504, 4626, \dodoi{10.1093/mnras/stab1109}

\bibitem[{{Kraljic} {et~al.}(2018){Kraljic}, {Arnouts}, {Pichon}, {Laigle}, {de la Torre}, {Vibert}, {Cadiou}, {Dubois}, {Treyer}, {Schimd}, {Codis}, {de Lapparent}, {Devriendt}, {Hwang}, {Le Borgne}, {Malavasi}, {Milliard}, {Musso}, {Pogosyan}, {Alpaslan}, {Bland-Hawthorn}, \& {Wright}}]{2018MNRAS.474..547K}
{Kraljic}, K., {Arnouts}, S., {Pichon}, C., {et~al.} 2018, \mnras, 474, 547, \dodoi{10.1093/mnras/stx2638}

\bibitem[{{Kraljic} {et~al.}(2019){Kraljic}, {Pichon}, {Dubois}, {Codis}, {Cadiou}, {Devriendt}, {Musso}, {Welker}, {Arnouts}, {Hwang}, {Laigle}, {Peirani}, {Slyz}, {Treyer}, \& {Vibert}}]{2019MNRAS.483.3227K}
{Kraljic}, K., {Pichon}, C., {Dubois}, Y., {et~al.} 2019, \mnras, 483, 3227, \dodoi{10.1093/mnras/sty3216}

\bibitem[{{Krolewski} {et~al.}(2019){Krolewski}, {Ho}, {Chen}, {Chan}, {Tenneti}, {Bizyaev}, \& {Kraljic}}]{2019ApJ...876...52K}
{Krolewski}, A., {Ho}, S., {Chen}, Y.-C., {et~al.} 2019, \apj, 876, 52, \dodoi{10.3847/1538-4357/ab1010}

\bibitem[{{Kugel} \& {van de Weygaert}(2024)}]{2024arXiv240716489K}
{Kugel}, R., \& {van de Weygaert}, R. 2024, arXiv e-prints, arXiv:2407.16489, \dodoi{10.48550/arXiv.2407.16489}

\bibitem[{{Laigle} {et~al.}(2015){Laigle}, {Pichon}, {Codis}, {Dubois}, {Le Borgne}, {Pogosyan}, {Devriendt}, {Peirani}, {Prunet}, {Rouberol}, {Slyz}, \& {Sousbie}}]{2015MNRAS.446.2744L}
{Laigle}, C., {Pichon}, C., {Codis}, S., {et~al.} 2015, \mnras, 446, 2744, \dodoi{10.1093/mnras/stu2289}

\bibitem[{{Laigle} {et~al.}(2018){Laigle}, {Pichon}, {Arnouts}, {McCracken}, {Dubois}, {Devriendt}, {Slyz}, {Le Borgne}, {Benoit-L{\'e}vy}, {Hwang}, {Ilbert}, {Kraljic}, {Malavasi}, {Park}, \& {Vibert}}]{2018MNRAS.474.5437L}
{Laigle}, C., {Pichon}, C., {Arnouts}, S., {et~al.} 2018, \mnras, 474, 5437, \dodoi{10.1093/mnras/stx3055}

\bibitem[{Lee \& Erdogdu(2007)}]{Lee2007}
Lee, J., \& Erdogdu, P. 2007, The Astrophysical Journal, 671, 1248, \dodoi{10.1086/523351}

\bibitem[{{Lee} \& {Pen}(2000)}]{2000ApJ...532L...5L}
{Lee}, J., \& {Pen}, U.-L. 2000, \apjl, 532, L5, \dodoi{10.1086/312556}

\bibitem[{{Libeskind} {et~al.}(2013){Libeskind}, {Hoffman}, {Forero-Romero}, {Gottl{\"o}ber}, {Knebe}, {Steinmetz}, \& {Klypin}}]{2013MNRAS.428.2489L}
{Libeskind}, N.~I., {Hoffman}, Y., {Forero-Romero}, J., {et~al.} 2013, \mnras, 428, 2489, \dodoi{10.1093/mnras/sts216}

\bibitem[{{Libeskind} {et~al.}(2014){Libeskind}, {Knebe}, {Hoffman}, \& {Gottl{\"o}ber}}]{2014MNRAS.443.1274L}
{Libeskind}, N.~I., {Knebe}, A., {Hoffman}, Y., \& {Gottl{\"o}ber}, S. 2014, \mnras, 443, 1274, \dodoi{10.1093/mnras/stu1216}

\bibitem[{{L{\'o}pez}(2024)}]{2024PASP..136c7001L}
{L{\'o}pez}, P. 2024, \pasp, 136, 037001, \dodoi{10.1088/1538-3873/ad31c9}

\bibitem[{{L{\'o}pez} {et~al.}(2021){L{\'o}pez}, {Cautun}, {Paz}, {Merch{\'a}n}, \& {van de Weygaert}}]{2021MNRAS.502.5528L}
{L{\'o}pez}, P., {Cautun}, M., {Paz}, D., {Merch{\'a}n}, M., \& {van de Weygaert}, R. 2021, \mnras, 502, 5528, \dodoi{10.1093/mnras/stab451}

\bibitem[{{L{\'o}pez} {et~al.}(2025){L{\'o}pez}, {van de Weygaert}, \& {Merch{\'a}n}}]{2025arXiv250501298L}
{L{\'o}pez}, P., {van de Weygaert}, R., \& {Merch{\'a}n}, M. 2025, arXiv e-prints, arXiv:2505.01298, \dodoi{10.48550/arXiv.2505.01298}

\bibitem[{{Neyrinck} {et~al.}(2020){Neyrinck}, {Aragon-Calvo}, {Falck}, {Szalay}, \& {Wang}}]{2020OJAp....3E...3N}
{Neyrinck}, M., {Aragon-Calvo}, M.~A., {Falck}, B., {Szalay}, A.~S., \& {Wang}, J. 2020, The Open Journal of Astrophysics, 3, 3, \dodoi{10.21105/astro.1904.03201}

\bibitem[{Pahwa {et~al.}(2016)Pahwa, Libeskind, Tempel, Hoffman, Tully, Courtois, Gottlöber, Steinmetz, \& Sorce}]{2016MNRAS.457..695P}
Pahwa, I., Libeskind, N.~I., Tempel, E., {et~al.} 2016, Monthly Notices of the Royal Astronomical Society, 457, 695, \dodoi{10.1093/mnras/stv2930}

\bibitem[{{Peebles}(1969)}]{1969ApJ...155..393P}
{Peebles}, P.~J.~E. 1969, \apj, 155, 393, \dodoi{10.1086/149876}

\bibitem[{{Porciani} {et~al.}(2002{\natexlab{a}}){Porciani}, {Dekel}, \& {Hoffman}}]{2002MNRAS.332..325P}
{Porciani}, C., {Dekel}, A., \& {Hoffman}, Y. 2002{\natexlab{a}}, \mnras, 332, 325, \dodoi{10.1046/j.1365-8711.2002.05305.x}

\bibitem[{{Porciani} {et~al.}(2002{\natexlab{b}}){Porciani}, {Dekel}, \& {Hoffman}}]{2002MNRAS.332..339P}
---. 2002{\natexlab{b}}, \mnras, 332, 339, \dodoi{10.1046/j.1365-8711.2002.05306.x}

\bibitem[{{Sch{\"a}fer}(2009)}]{2009IJMPD..18..173S}
{Sch{\"a}fer}, B.~M. 2009, International Journal of Modern Physics D, 18, 173, \dodoi{10.1142/S0218271809014388}

\bibitem[{{Sheng} {et~al.}(2022){Sheng}, {Li}, {Yu}, {Wang}, {Wang}, \& {Kang}}]{2022PhRvD.105f3540S}
{Sheng}, M.-J., {Li}, S., {Yu}, H.-R., {et~al.} 2022, \prd, 105, 063540, \dodoi{10.1103/PhysRevD.105.063540}

\bibitem[{{Sheng} {et~al.}(2023){Sheng}, {Yu}, {Li}, {Liao}, {Du}, {Wang}, {Wang}, {Xu}, {Genel}, \& {Irodotou}}]{2023ApJ...943..128S}
{Sheng}, M.-J., {Yu}, H.-R., {Li}, S., {et~al.} 2023, \apj, 943, 128, \dodoi{10.3847/1538-4357/acae92}

\bibitem[{{Shi} {et~al.}(2015){Shi}, {Wang}, \& {Mo}}]{2015ApJ...807...37S}
{Shi}, J., {Wang}, H., \& {Mo}, H.~J. 2015, \apj, 807, 37, \dodoi{10.1088/0004-637X/807/1/37}

\bibitem[{Stoica {et~al.}(2007)Stoica, Martínez, \& Saar}]{2007JRSSC..56....1S}
Stoica, R.~S., Martínez, V.~J., \& Saar, E. 2007, Journal of the Royal Statistical Society: Series C (Applied Statistics) 56 (4, 56, 1, \dodoi{10.1111/j.1467-9876.2007.00587.x}

\bibitem[{{Tang} {et~al.}(2025){Tang}, {Wang}, {Wang}, {Sheng}, {Yu}, \& {Xu}}]{2025ApJ...982..197T}
{Tang}, X.-x., {Wang}, P., {Wang}, W., {et~al.} 2025, \apj, 982, 197, \dodoi{10.3847/1538-4357/adbbd7}

\bibitem[{{Tempel} \& {Libeskind}(2013)}]{2013ApJ...775L..42T}
{Tempel}, E., \& {Libeskind}, N.~I. 2013, \apjl, 775, L42, \dodoi{10.1088/2041-8205/775/2/L42}

\bibitem[{Tempel \& Libeskind(2013)}]{Tempel-2013}
Tempel, E., \& Libeskind, N.~I. 2013, The Astrophysical Journal, 775, L42, \dodoi{10.1088/2041-8205/775/2/l42}

\bibitem[{{Tempel} {et~al.}(2016){Tempel}, {Stoica}, {Kipper}, \& {Saar}}]{2016A&C....16...17T}
{Tempel}, E., {Stoica}, R.~S., {Kipper}, R., \& {Saar}, E. 2016, Astronomy and Computing, 16, 17, \dodoi{10.1016/j.ascom.2016.03.004}

\bibitem[{{Tempel} {et~al.}(2013){Tempel}, {Stoica}, \& {Saar}}]{2013MNRAS.428.1827T}
{Tempel}, E., {Stoica}, R.~S., \& {Saar}, E. 2013, \mnras, 428, 1827, \dodoi{10.1093/mnras/sts162}

\bibitem[{{van de Weygaert} \& {Bertschinger}(1996)}]{1996MNRAS.281...84V}
{van de Weygaert}, R., \& {Bertschinger}, E. 1996, \mnras, 281, 84, \dodoi{10.1093/mnras/281.1.84}

\bibitem[{{van de Weygaert} \& {Bond}(2008)}]{2008LNP...740..335V}
{van de Weygaert}, R., \& {Bond}, J.~R. 2008, in A Pan-Chromatic View of Clusters of Galaxies and the Large-Scale Structure, ed. M.~{Plionis}, O.~{L{\'o}pez-Cruz}, \& D.~{Hughes}, Vol. 740, 335, \dodoi{10.1007/978-1-4020-6941-3_10}

\bibitem[{{van Haarlem} \& {van de Weygaert}(1993)}]{1993ApJ...418..544V}
{van Haarlem}, M., \& {van de Weygaert}, R. 1993, \apj, 418, 544, \dodoi{10.1086/173416}

\bibitem[{Varela {et~al.}(2012)Varela, Betancort-Rijo, Trujillo, \& Ricciardelli}]{2012ApJ...744...82V}
Varela, J., Betancort-Rijo, J., Trujillo, I., \& Ricciardelli, E. 2012, The Astrophysical Journal, 744, 82, \dodoi{10.1088/0004-637x/744/2/82}

\bibitem[{Wake {et~al.}(2017)Wake, Bundy, Diamond-Stanic, Yan, Blanton, Bershady, Sánchez-Gallego, Drory, Jones, Kauffmann, Law, Li, MacDonald, Masters, Thomas, Tinker, Weijmans, \& Brownstein}]{MaNGA-sample}
Wake, D.~A., Bundy, K., Diamond-Stanic, A.~M., {et~al.} 2017, The Astronomical Journal, 154, 86, \dodoi{10.3847/1538-3881/aa7ecc}

\bibitem[{Walmsley {et~al.}(2022)Walmsley, Lintott, Géron, Kruk, Krawczyk, Willett, Bamford, Kelvin, Fortson, Gal, Keel, Masters, Mehta, Simmons, Smethurst, Smith, Baeten, \& Macmillan}]{Galaxy-Zoo-DECaLS}
Walmsley, M., Lintott, C., Géron, T., {et~al.} 2022, Monthly Notices of the Royal Astronomical Society, 509, 3966, \dodoi{10.1093/mnras/stab2093}

\bibitem[{Wang {et~al.}(2018)Wang, Guo, Kang, \& Libeskind}]{2018ApJ...866..138W}
Wang, P., Guo, Q., Kang, X., \& Libeskind, N.~I. 2018, The Astrophysical Journal, 866, 138, \dodoi{10.3847/1538-4357/aae20f}

\bibitem[{Wang \& Kang(2018)}]{2018MNRAS.473.1562W}
Wang, P., \& Kang, X. 2018, Monthly Notices of the Royal Astronomical Society, 473, 1562, \dodoi{10.1093/mnras/stx2466}

\bibitem[{Wang {et~al.}(2021)Wang, Libeskind, Tempel, Kang, \& Guo}]{Wang2021NA}
Wang, P., Libeskind, N.~I., Tempel, E., Kang, X., \& Guo, Q. 2021, Nature Astronomy, 5, 839, \dodoi{10.1038/s41550-021-01380-6}

\bibitem[{Wang {et~al.}(2020)Wang, Libeskind, Tempel, Pawlowski, Kang, \& Guo}]{2020ApJ...900..129W}
Wang, P., Libeskind, N.~I., Tempel, E., {et~al.} 2020, The Astrophysical Journal, 900, 129, \dodoi{10.3847/1538-4357/aba6ea}

\bibitem[{{Wang} {et~al.}(2025){Wang}, {Tang}, {Wang}, {Libeskind}, {Tempel}, {Wang}, {Zhang}, {Sheng}, {Yu}, \& {Xu}}]{2025ApJ...983..100W}
{Wang}, P., {Tang}, X.-x., {Wang}, H.-d., {et~al.} 2025, \apj, 983, 100, \dodoi{10.3847/1538-4357/adc0a2}

\bibitem[{{Wang} {et~al.}(2024){Wang}, {Wang}, {Guo}, {Kang}, {Libeskind}, {Gal{\'a}rraga-Espinosa}, {Springel}, {Kannan}, {Hernquist}, {Pakmor}, {Yu}, {Bose}, {Guo}, {Yu}, \& {Hern{\'a}ndez-Aguayo}}]{Wangwei2024}
{Wang}, W., {Wang}, P., {Guo}, H., {et~al.} 2024, \mnras, 532, 4604, \dodoi{10.1093/mnras/stae1801}

\bibitem[{Welker {et~al.}(2014)Welker, Devriendt, Dubois, Pichon, \& Peirani}]{2014MNRAS.445L..46W}
Welker, C., Devriendt, J., Dubois, Y., Pichon, C., \& Peirani, S. 2014, Monthly Notices of the Royal Astronomical Society, 445, L46, \dodoi{10.1093/mnrasl/slu106}

\bibitem[{Welker {et~al.}(2020)Welker, Bland-Hawthorn, van~de Sande, Lagos, Elahi, Obreschkow, Bryant, Pichon, Cortese, Richards, Croom, Goodwin, Lawrence, Sweet, Lopez-Sanchez, Medling, Owers, Dubois, \& Devriendt}]{2020MNRAS.491.2864W}
Welker, C., Bland-Hawthorn, J., van~de Sande, J., {et~al.} 2020, Monthly Notices of the Royal Astronomical Society, 491, 2864, \dodoi{10.1093/mnras/stz2860}

\bibitem[{Westfall {et~al.}(2019)Westfall, Cappellari, Bershady, Bundy, Belfiore, Ji, Law, Schaefer, Shetty, Tremonti, Yan, Andrews, Brownstein, Cherinka, Coccato, Drory, Maraston, Parikh, Sánchez-Gallego, Thomas, Weijmans, Barrera-Ballesteros, Du, Goddard, Li, Masters, Ibarra~Medel, Sánchez, Yang, Zheng, \& Zhou}]{MaNGA-pipeline}
Westfall, K.~B., Cappellari, M., Bershady, M.~A., {et~al.} 2019, The Astronomical Journal, 158, 231, \dodoi{10.3847/1538-3881/ab44a2}

\bibitem[{{White}(1984)}]{1984ApJ...286...38W}
{White}, S.~D.~M. 1984, \apj, 286, 38, \dodoi{10.1086/162573}

\bibitem[{Willett {et~al.}(2013)Willett, Lintott, Bamford, Masters, Simmons, Casteels, Edmondson, Fortson, Kaviraj, Keel, Melvin, Nichol, Raddick, Schawinski, Simpson, Skibba, Smith, \& Thomas}]{Galaxy-zoo-2}
Willett, K.~W., Lintott, C.~J., Bamford, S.~P., {et~al.} 2013, Monthly Notices of the Royal Astronomical Society, 435, 2835, \dodoi{10.1093/mnras/stt1458}

\bibitem[{{Xia} {et~al.}(2021){Xia}, {Neyrinck}, {Cai}, \& {Arag{\'o}n-Calvo}}]{2021MNRAS.506.1059X}
{Xia}, Q., {Neyrinck}, M.~C., {Cai}, Y.-C., \& {Arag{\'o}n-Calvo}, M.~A. 2021, \mnras, 506, 1059, \dodoi{10.1093/mnras/stab1713}

\bibitem[{{Xu} {et~al.}(2022){Xu}, {Chen}, {Shi}, {Zhou}, {Bizyaev}, {Bao}, {Beom}, {Fern{\'a}ndez-Trincado}, \& {Cao}}]{2022MNRAS.511.4685X}
{Xu}, H., {Chen}, Y., {Shi}, Y., {et~al.} 2022, \mnras, 511, 4685, \dodoi{10.1093/mnras/stac354}

\bibitem[{{Yu} {et~al.}(2020){Yu}, {Motloch}, {Pen}, {Yu}, {Wang}, {Mo}, {Yang}, \& {Jing}}]{2020PhRvL.124j1302Y}
{Yu}, H.-R., {Motloch}, P., {Pen}, U.-L., {et~al.} 2020, \prl, 124, 101302, \dodoi{10.1103/PhysRevLett.124.101302}

\bibitem[{{Zhang} {et~al.}(2023){Zhang}, {Lee}, {Krolewski}, {Shi}, {Horowitz}, \& {Kooistra}}]{2023ApJ...954...49Z}
{Zhang}, B., {Lee}, K.-G., {Krolewski}, A., {et~al.} 2023, \apj, 954, 49, \dodoi{10.3847/1538-4357/ace695}

\bibitem[{{Zhang} {et~al.}(2025){Zhang}, {Yang}, {Guo}, {Wang}, \& {Shi}}]{zhang2025galaxyhalopropertiescosmic}
{Zhang}, Y., {Yang}, X., {Guo}, H., {Wang}, P., \& {Shi}, F. 2025, \mnras, 539, 1692, \dodoi{10.1093/mnras/staf611}

\bibitem[{{Zhang} {et~al.}(2015){Zhang}, {Yang}, {Wang}, {Wang}, {Luo}, {Mo}, \& {van den Bosch}}]{2015ApJ...798...17Z}
{Zhang}, Y., {Yang}, X., {Wang}, H., {et~al.} 2015, \apj, 798, 17, \dodoi{10.1088/0004-637X/798/1/17}

\bibitem[{Zhou {et~al.}(2022)Zhou, Chen, Shi, Bizyaev, Guo, Bao, Xu, Yu, \& Brownstein}]{2022MNRAS.515.5081Z}
Zhou, Y., Chen, Y., Shi, Y., {et~al.} 2022, Monthly Notices of the Royal Astronomical Society, 515, 5081, \dodoi{10.1093/mnras/stac2016}

\end{thebibliography}
\bibliographystyle{aasjournal}


 \end{CJK*}
\end{document}